\PassOptionsToPackage{hidelinks}{hyperref}

\documentclass[preprint]{vgtc}               

\graphicspath{{figures/}{pictures/}{images/}{./}} 

\usepackage{times}                     

\usepackage{tabu}                      
\usepackage{booktabs}                  
\usepackage{lipsum}                    
\usepackage{mwe}                       

\usepackage{mathptmx}                  

\onlineid{1080}

\vgtccategory{Research}

\vgtcinsertpkg

\ieeedoi{10.1109/ISMAR-Adjunct68609.2025.00267}

\title{ShadAR: LLM-driven shader generation to transform visual perception in Augmented Reality}

\author{Yanni Mei\thanks{e-mail: yanni.mei@tu-darmstadt.de} %
\and Samuel Wendt\thanks{e-mail:samuel.wendt@stud.tu-darmstadt.de} %
\and Florian Müller\thanks{e-mail:florian.mueller@tu-darmstadt.de} %
\and Jan Gugenheimer\thanks{e-mail:jan.gugenheimer@tu-darmstadt.de}}
\affiliation{\scriptsize TU Darmstadt}

\teaser{
  \centering
  \includegraphics[width=.9\linewidth]{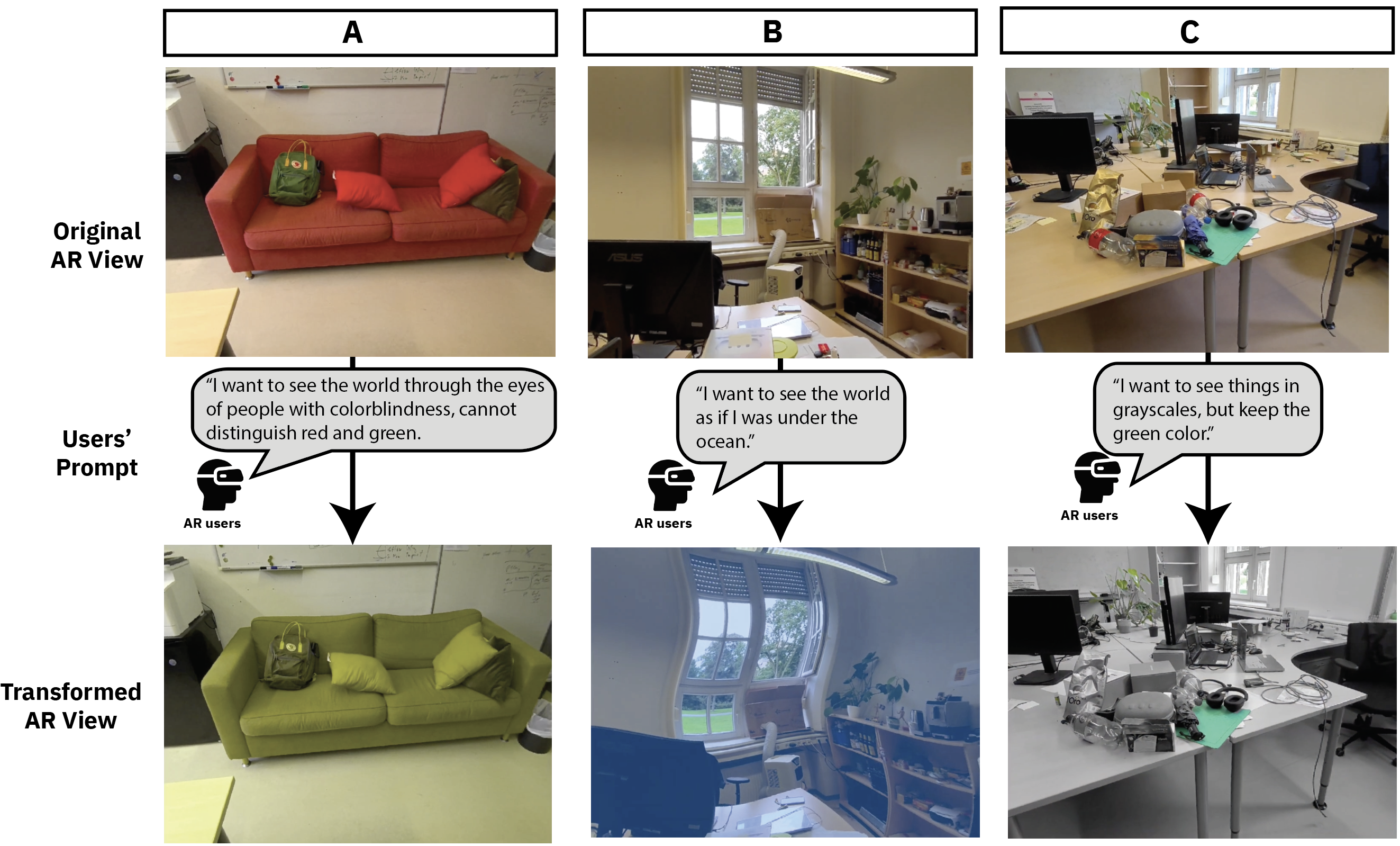}
  \caption{ShadAR enables AR users to transform their view in real time through natural language prompts, generating and applying custom HLSL shaders to AR viewport instantly. Examples: (A) simulate red–green colorblindness, (B) create an underwater visual effect, and (C) display the scene in grayscale while preserving green.}
  \label{fig:teaser}
}

\abstract{
Augmented Reality (AR) can simulate various visual perceptions, such as how individuals with colorblindness see the world. 
However, these simulations require developers to predefine each visual effect, limiting flexibility. 
We present ShadAR, an AR application enabling real-time transformation of visual perception through shader generation using large language models (LLMs). 
ShadAR allows users to express their visual intent via natural language, which is interpreted by an LLM to generate corresponding shader code. 
This shader is then compiled real-time to modify the AR headset’s viewport. 
We present our LLM-driven shader generation pipeline and demonstrate its ability to transform visual perception for inclusiveness and creativity.
} 


\begin{document}



\maketitle

\section{Introduction}

Augmented Reality (AR) has opened up new possibilities for directly modifying our visual perception, from simulating how others see the world, to creating entirely novel visual experience for artistic expressions\cite{langlotz2024design}. 
This capability has been explored in diverse domains. For example, it allows designers to understand the world as people with different visual capabilities (e.g, cataract) for more inclusive design decisions \cite{krosl2020cataract,zhang2022seeing,ates2015immersive}. It has also been used by artists to creatively stylize the environment, as perceived through the imagined eyes of Van Gogh\cite{geroimenko2014augmented}. 
However, current implementations of such visual perceptual simulations are typically created beforehand by the developer, designed for a single visual condition, and thus, lack flexibility for broader exploration.


Recent advances in large language models (LLMs) have demonstrated their ability to generate functional code from natural language instructions\cite{jiang2024survey}. Thus, LLM has be adopted in extended reality (XR) environment to support real-time XR content creation \cite{tang2025llm}. For example, LLMR enables users to generate or manipulate virtual objects in extended reality(XR) with voice command, using LLM to real-time generate CSharp code \cite{de2024llmr}.

While prior work has centered on generating CSharp code to create/manipulate XR objects, we take a different step: using LLMs to generate HLSL shaders for creating/changing AR visual experiences. In this project, we present ShadAR, a LLM-enabled shader generation system for augmented reality(AR), deployed on Meta Quest 3. ShadAR enables AR users to verbally express their desired visual experience (e.g., ``I want to see the world through the eyes of someone who is colorblind.''). Then, a \textit{ShaderGeneratorLLM}, guided by designed prompts, interprets the user's spoken command and generates HLSL shader code in real time. The shader is then compiled and immediately applied to the headset’s camera passthrough, transforming the user’s visual experience on the fly.

We show example applications to demonstrate ShadAR’s ability to flexibly interpret user intent and generate corresponding visual effects --- though not always perceptually accurate, these effects allow for rapid exploration of altered visual perceptions.

\section{Technical Architecture}

\begin{figure}
    \centering
    \includegraphics[width=1\linewidth]{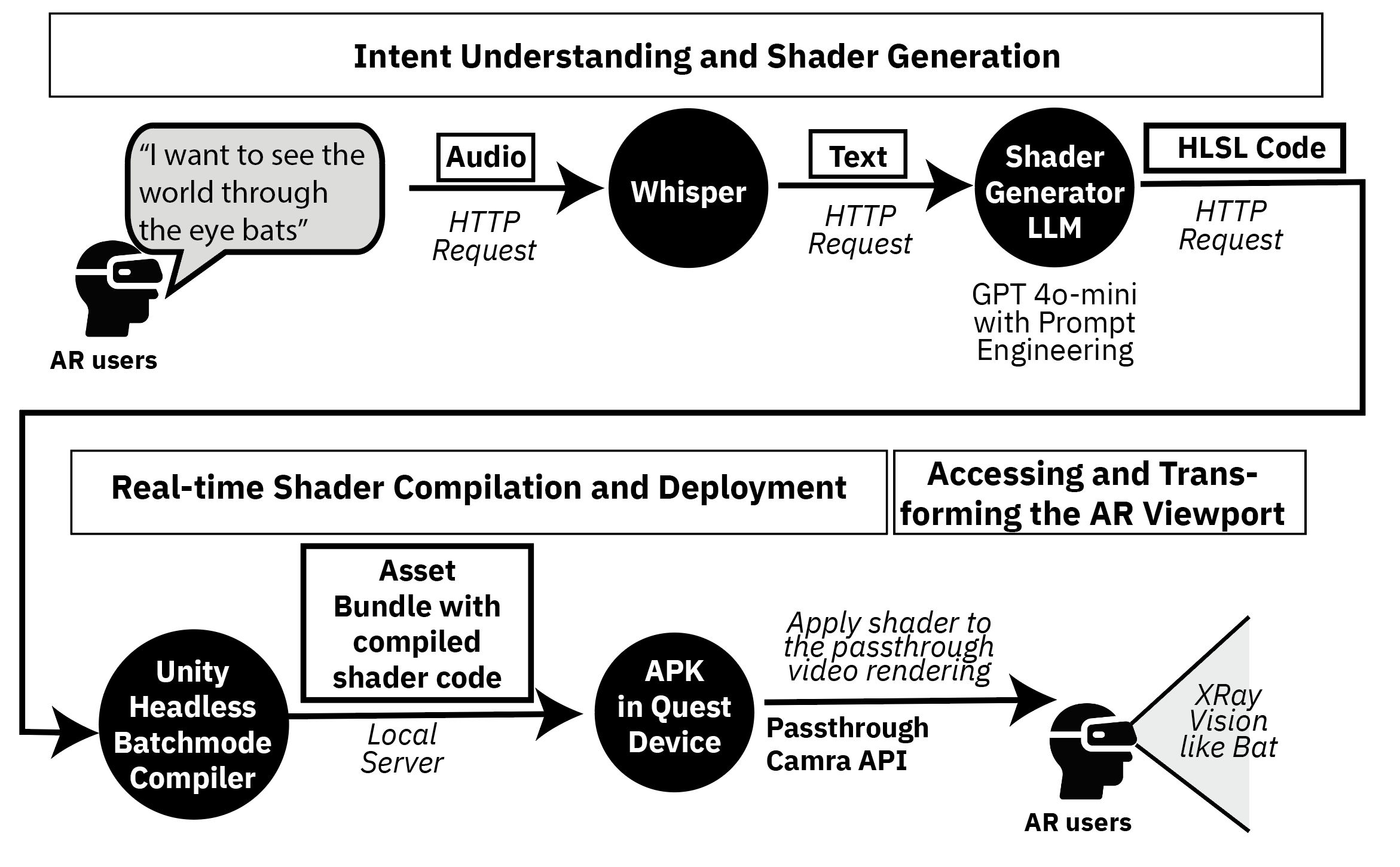}
    \caption{Technical architecture of ShadAR}
    \label{fig:architecture}
\end{figure}

To enable real-time visual transformations from natural language input, ShadAR integrates a pipeline (Fig.\ref{fig:architecture}) including three parts:

\textbf{Intent Understanding and Shader Generation.}
User voice commands are transcribed using Whisper and sent via HTTP to a prompt-engineered LLM (OpenAI o3-mini). The prompt incorporates domain knowledge (e.g., animal vision effects), example shader templates, and common pitfalls in shader programming to enhance output reliability.

\textbf{Real-time Shader Compilation and Deployment.} 
The generated shader code is compiled on a desktop using Unity in headless batch mode, packaged into an asset bundle, then deployed to the Quest device via a local web server. 

\textbf{Accessing and Transforming the AR Viewport.}
We utilize the \textit{Meta Passthrough Camera API} to access and change the AR headset’s video stream as a \textit{WebCamTexture} in \textit{Unity3D}. 

\section{Demo Experience}

\textbf{Verbally Prompt Visual Perceptions.} To experience ShadAR, the user needs to wear the Meta Quest 3 headset, and speak aloud the visual effect they want to experience—such as saying, \textit{``I want to see the world with heat vision.''} While describing their desired effect, they press and hold the Oculus B button to record their voice. After approximately 30 to 45 seconds, a custom shader is generated and applied to the video passthrough in the user’s AR viewport. Users can choose to save the shader for future use, or record a new voice command to generate new visual effect.

\textbf{Monoscopic Simulates Stereoscopic.} In our setup,  the AR viewport is rendered on a 2D quad displaying the left-eye camera feed, resulting in a monoscopic view. Then the compiled shader is applied to the material of the quad rendering the passthrough video, enabling real-time visual perception transformation.  While this is not a stereoscopic view, by carefully aligning the quad in 3D space --- at an appropriate distance and scale to the user's eye, we try to achieve a visually coherent and sufficiently immersive AR experience.

\section{Example Shader Generated by ShadAR}



In Fig.~\ref{fig:teaser}, we illustrate examples of using shader generation for visual perception transformation. The full video can be accessed via this link \url{https://youtu.be/XS90nrJnel4}. These examples span a range of goals:

\textbf{Accessibility} Simulation of colorblindness (Fig.~\ref{fig:teaser}-A) or other visual impairments can help designers understand the challenges faced by people with different visual capabilities, enabling more inclusive design choices.

\textbf{Creative Expression} Creating effects like viewing the world from under the ocean (Fig.~\ref{fig:teaser}-B) showcases how such transformations enable artists to explore unique aesthetics and convey specific narratives.

\textbf{Target Highlighting} Prompts like \textit{``I want to see things in grayscale except green''} make it easier for users to find the green folder in cluttered environments (Fig.~\ref{fig:teaser}-C). It illustrates how visual perception transformations can support navigation tasks in productivity scenarios.

\section{Conclusion}
In this work, we present ShadAR, an AR application that leverages LLMs to interpret users’ verbal commands and generate corresponding shader code, enabling real-time transformation of visual perception.
We explain the system architecture and showcase examples of generated visual experiences that support inclusiveness, creativity, and searching.
We hope to make AR a more flexible ``visual experience machine'', utilizing the code generation capability of LLMs.

\acknowledgments{ This project has been funded by the LOEWE initiative (Hesse, Germany) within the
emergenCITY center.
}

\bibliographystyle{abbrv-doi}

\bibliography{template}
\end{document}